# Hybrid MoS$_2$-gap-mode metasurfaces photodetectors

**Peinan Ni[1], Andrés De Luna Bugallo[2,*], Xun Yang[3], Victor M. Arellano Arreola[2], Mario Flores Salazar[2], Elodie Strupiechonski[2], Blandine Alloing[1], Chongxin Shan[3], and Patrice Genevet[1,*]**

[1]*Université Cote d'Azur, CNRS, CRHEA, rue Bernard Gregory, Sophia Antipolis 06560 Valbonne, France*

[2] *CONACYT- Cinvestav Unidad, Querétaro, Querétaro, Qro 76230, Mexico*

[3]*School of Physics and Engineering, Zhengzhou University, Zhengzhou 450052, China*

[*] *corresponding authors: aluna@cinvestav.mx and Patrice.Genevet@crhea.cnrs.fr*

Two-dimensional molybdenum disulfide (MoS$_2$) featuring atomically thin thickness and unique electronic structure with favorable bandgap has been widely recognized as an attractive new material for the development of the next generation of ultra-compact, light-weight optoelectronic components. In parallel, the recently emerged metasurfaces have demonstrated exceptional controllability over electromagnetic field within ultra-compact subwavelength dimension offering an unprecedented approach to improve the performance of optoelectronic devices. In this work, we are proposing an integration of metasurfaces with 2D semiconductor materials to achieve polarization sensitive, fast-response photodetectors. The reported devices are among the most compact hybrid MoS$_2$-gap-plasmon metasurface detectors. Relying on the significant electromagnetic field confinement provided by the metasurfaces to enhance light absorption and to reduce the surface states, which generally limit the photo-generated carriers lifetime, we measured enhanced photocurrent and a fast detection speed. Moreover, the strong optical anisotropy introduced by the metasurfaces is used to efficiently control the polarization sensitivity of the photodetector. This work provides a feasible and effective solution to improve the performance of two-dimensional materials based photodetectors.

**KEYWORDS**: Gap plasmon polariton; metasurfaces; 2D materials; MoS2; photodetector

**Introduction**

The unique optical and optoelectronic properties of atomically thin and layered two dimensional (2D) transition metal dichalcogenides (TMDCs) have been exploited to study and demonstrate a great variety of emerging optoelectronic nano-devices in the visible spectral range.[1-3] Molybdenum disulfide ($MoS_2$), a typical 2D TMDC, has attracted extensive attention as a promising candidate for the next-generation ultra-compact photodetectors due to its excellent electronic and optical properties, *i.e.* a direct bandgap of 1.8 eV in monolayer and an indirect bandgap of 1.3 eV for bulk or multilayer, high carrier mobility of about 200 $cm^2V^{-1}s^{-1}$ for monolayer and about 500 $cm^2V^{-1}s^{-1}$ for few layers, wide wavelength absorption (350-1000 nm) and high power conversion efficiencies.[4-6] However, currently the performances of the $MoS_2$ based photodetectors are mainly limited, essentially due to their extreme thinness, which leads to a decrease of light-matter interaction and therefore weak light absorption, and also because of their large surface to volume ratio, which not only makes them very sensitive to the environment but also largely limits their response speed.[7,8] Indeed, the large amount of surface trapping states of $MoS_2$ can capture photo-generated electrons and results in the long lifetimes of photo-generated holes, leading to slow photoresponse generally ranging from several microseconds to seconds.[9-10] Great efforts have thus been devoted to improve the performance of $MoS_2$ based photodetectors by solving those problems. For example, the localized surface plasmon of metallic nanostructures has been employed to enhance the light absorption of $MoS_2$.[11] In addition, both vacuum annealing and surface encapsulation approaches have been utilized to improve the surface quality of $MoS_2$-based devices by reducing the trapping states.[9,10] Emerging as a rapidly developing technique, metasurfaces, a new class of 2D optical components have been demonstrated, showing extraordinary capabilities for molding the light properties in a very compact and efficient way. Metasurfaces are thus giving rise to a large variety of novel optical components including frequency selective surface (FSS), polarization converter, wavefront shaping, hologram, hybrid and free-form metaoptics.[12-17] This compelling features characterize metasurfaces as versatile and indispensable platforms for spectral and spatial manipulation of electromagnetic waves within subwavelength regime. Extending the remarkable controllability of metasurfaces into 2D materials-based optoelectronic devices through hybridization and integrations will open up great opportunities to realize the high performance ultra-compact optoelectronic devices.

In this work, high performance $MoS_2$ photodetectors with polarization sensitivity and fast response are demonstrated through the integration of a gap-surface-plasmon (GSP) metasurface. The resonance mode of the GSP metasurface is designed to couple with the neutral A-exciton resonance of monolayer $MoS_2$. By doing this, the light-matter interactions inside the monolayer $MoS_2$ are selectively enhanced due to the strong light confinement of the GSP metasurface. Meanwhile, placing GSP in close proximity to form a metasurface, i.e. by placing nanostripes parallel and close to each other's, we form an ensemble of adjacent nanostripes that serve as antenna-assisted electrodes to collect the photo-generated carriers very efficiently.[18] On the other hand, since the electromagnetic field inside this structure is concentrated within the small volume of the metasurface, the proposed photodetectors are less affected by the large amount of surface states located on the entire surface and, as a result, show a fast photo-response. Furthermore, we show that the strong optical anisotropy introduced by the metasurface can be used to effectively control the polarization sensitivity of the photodetectors.

**Results and discussion**

Since 2D materials present no dangling bonds on the surfaces they are perfect candidates to implement integrated hybrid structure with other materials beyond the limitation of lattice matching. In particular, the integration of a plasmonic structure can remarkably enhance the light-matter interactions inside a 2D material[19-21]. To this end, the hybrid MoS$_2$-gap-mode metasurface photodetector is designed and realized by integrating metallic nanostripe-arrays onto the MoS$_2$ flake, as shown in Fig. 1. This integration can greatly improve the performance of MoS$_2$ based photodetectors by making advantage of the extraordinary capabilities of metasurface in manipulating light-matter interactions at subwavelength dimensions. In the proposed structure, each metallic Au nanostripe acts as individual gap plasmon cavities. The gap plasmon resonance established inside this structure is determined by the standard Fabry-Pérot resonance formula:

$$wk_0 n_{gsp} + \phi = m\pi \qquad \textbf{\textit{Eq. 1}}$$

where $w$ is the width of the nanostripe, $k_0$ is the vacuum wave number, $n_{\text{gsp}}$ is effective index of the GSP, m is an integer defining the mode order, and $\varphi$ is an additional phase shift acquired by the gap plasmon upon reflection at the edges of the metallic top strip. It can be seen that the width of each nanostripe plays a critical role in the design to determine the resonance conditions of the GSP metasurface. Surface plasmon excited and propagating back and forth across the nanostripe acquires propagation and reflection phase delays leading to a resonant behavior, as discussed in Eq.1. The additional reflection phase shift depends on the structural and materials parameters and increases with increasing gap layer to $\phi \sim \frac{3}{5}\pi$ for a relatively thick cladding layer of about 100 nm in this structure. Using Eq1, we estimate that the effective mode index of the GSP mode is $n_{gsp} \sim 1.8$.

The GSP resonance of the metasurface integrated in this structure is designed to match the exciton A resonance of the MoS$_2$ monolayer at around 650 nm. For such purpose, finite difference time domain (FDTD) simulations are performed to investigate the resonance of the bare GSP metasurface to determine the geometry of the nanostripe arrays. Figure 2(a) shows the simulated reflectance of the designed GSP metasurface as a function of the width of the Au nanostripe while the period of the stripe array is fixed at 445 nm. At resonance, the optical energy can be effectively coupled into the GSP mode. Meanwhile, due to the close proximity of the nanostripes and bottom metallic mirror, the electromagnetic field is tightly confined, mostly within the gap region, in a mode volume in the order of $V_{eff} \sim 3.5\ 10^{-4}\ \mu m^3$, almost two order of magnitude smaller than the diffraction limited mode volume $\left(\frac{\lambda}{2n_{gsp}}\right)^3$ considering $\lambda$ as the free space PL wavelength, as confirmed by the full electromagnetic wave simulations shown in Fig. 2(b).

The large confinement factor provided by the plasmonic nanocavities can greatly enhance the photocurrent response of the 2D materials-based photodetectors by improving the light absorption.[18] To benefit from the electromagnetic field confinement enabled by the proposed GSP metasurface in this work, the MoS$_2$ monolayer is placed in close contact with the metallic nanostripes within the large electromagnetic field enhancement regions, as illustrated in Fig. 2(b). Figure 3 shows the photoluminescence (PL) spectra of the MoS$_2$ flakes both inside and outside the GSP metasurface. PL enhancement can be clearly observed from the MoS$_2$ flake inside the metasurface as a result of the tight electromagnetic confinement of the GSP metasurfaces, which is in good agreement with the design of the GSP mode coupling with the exciton resonance of MoS$_2$. It is also worth pointing out that the large PL enhancement is observed even if the PL emission of the MoS$_2$ is partially quenched by the absorption of the Au stripes (considering that the MoS$_2$ flake is in direct contact with top Au stripes). In addition to the strong enhancement of light-matter interactions, the GSP metasurface will also act as electrodes which greatly increases the collection efficiency of the photo-generated carriers, further improving the performance of the MoS$_2$ based photodetectors. Due to the relatively small electron and holes mobility ($\mu$) in MoS2, around $\mu_e = 150 \text{ cm}^2 \text{ V}^{-1} \text{ S}^{-1}$ and $\mu_h = 80 \text{ cm}^2 \text{ V}^{-1} \text{ S}^{-1}$ respectively, and the short carrier lifetime ($\tau$) due to various recombination and phonon excitations $\tau^{(MoS2)} \approx 300 \text{ ps}$, the diffusion lengths ( $L = \sqrt{\frac{\mu_h k_b T}{e} \tau}$ ) with of photogenerated electrons and holes are in the order of $L_e^{(Mos_2)} \approx 300nm$ and $L_h^{(Mos2)} = 200nm$.[21] We thus designed nanosize gap between adjacent collecting antennas of 370nm, in agreement with the photogenerated carrier diffusion length to maximize photoresponse. To further investigate the impact on the crystal structure of the MoS$_2$ flakes once the photodetector was fabricated, we performed Raman spectroscopy inside and outside the metallic nanostripes, as shown in Fig. 3(b). The A1g peak is located at 405 cm$^{-1}$ for both cases while the E12g is centered at 384.54 cm$^{-1}$ outside and 383.3 cm-$^1$ inside the gold grating. The slight Raman downshift of the E12g peak suggest an induced strain by the metallic structures, however, since A1g peak position remains unchanged one can dismiss charge transfer from the plasmonic cavity to the underlying MoS$_2$ flakes. These results are in good agreement with the high photoluminescence emission intensity observed in Fig. 3(a).

Furthermore, the remarkable near-field electromagnetic enhancement of plasmonic nanostructures has demonstrated exceptional polarization manipulation capability. For example, the combination of chiral plasmonic fields with the valley-selective response of TMD materials has been successfully employed to increase the valley polarization and direct valley-selective exciton emissions.[22-24] In this work, we show that the strong anisotropy introduced by the GSP metasurface design provides a feasible and effective approach to control the polarization sensitivity of the photodetector. It is found that the photocurrent of the structure is at a maximum when the electric field component of the incident light is perpendicular to the Au stripes (defined as 0° polarization) and at a minimum when the electric field component is parallel to the Au stripes (defined as 90° polarization), as shown in Fig. 4(a), which gives rise to an about 30-fold polarization sensitivity. The polarization sensitivity of the proposed structure can be further characterized by the polarization ratio $\rho = (I_{TM} - I_{TE})/(I_{TM} + I_{TE})$, leading to a polarization ratio of ~0.94. Note that the anisotropic crystal structure of 2D materials can also result in directional dependence of the light absorption and the corresponding polarization-sensitive photocurrent generation, which has been demonstrated successfully to realize polarization-sensitive photodetectors. However, the reported polarization sensitive photodetectors relying on this effect only exhibit relatively small polarization ratio. For example, a ratio about 0.5 is obtained from a photodetector using a black phosphorus vertical p–n junction,[25] 0.655 from a photodetector based on a $MoS_2$/GaAs heterojunction[26] and 0.709 from a photodetector based on black phosphorus on $WSe_2$ photogate vertical heterostructure.[27] On the other hand, since the orientation of the Au stripes has not been deliberately aligned with respect to the $MoS_2$ in our structure, the polarization sensitivity is mainly due to the large anisotropy introduced by the GSP metasurface. In contrast, the larger polarization ratio observed from this structure compared to the previous works demonstrates its superior advantage to manipulate the polarization sensitivity of photodetectors. Moreover, our work also suggests that other metasurface designs such as zigzag wires, asymmetric metallic rods or chiral structures could also be employed for polarization sensitive photodetections.

Finally, the photocurrent of the proposed hybrid structure shows a linear dependence on the applied voltage, as shown in Fig. 4(b). In addition, the response speed is another important figure of merit of a photodetector, which determines the ability of the detector to follow the fast varying optical signals. The MoS2 based photodetectors usually show a slow response in the range from several microseconds to seconds due to the large amount of surface trapping states, which capture the photo-generated carriers and release them slowly when the incident light is turned off. The tightly confined electromagnetic fields inside the hybrid structure within a small mode volume greatly reduce the influence of surface states on the generation of photocurrent and as a result increases its photodetection response speed. To investigate the response speed of the hybrid $MoS_2$ gap-plasmon metasurface photodetector, a 650 nm laser modulated by an optical chopper with tunable frequencies was used as the incident light source. The photocurrent measured at different chopper frequency is shown in Fig. 4(c). It can be seen that the structure can maintain large photocurrent at higher light switching frequency with only a slight loss in performance, indicating that it is capable of detecting fast varying optical signals.

**Experimental methods**

$MoS_2$ flakes were grown on $SiO_2$/Si substrates in a quartz-tube-in-a-furnace vapor-phased deposition system at ambient pressure in a quartz tube with argon as a carrier gas. $MoO_3$ (99% Sigma Aldrich) and Sulfur powders (99% Sigma Aldrich) are used as precursors, while $MoO_3$ is placed at the center of the furnace in an alumina boat, sulfur is situated upstream at the edge of the furnace. The temperature of the system is first increased from room temperature to 300°C with a rate of 10°C/min, after 30 minutes the temperature is raised up to 700°C with a ramp of 30°C/min, the system is immediately cooled down naturally once it reaches the growth temperature (700°C). To fabricate the hybrid $MoS_2$-gap plasmon metasurface structure, 5 nm Ti/ 200 nm Au were first deposited onto 1cm x 1cm sapphire substrates by electron beam evaporator acting as the metallic mirror. After that, a 100 nm $SiO_2$ dielectric layer was deposited by a conventional sputtering system. Then, an optical lithography step is performed to define large metallic contacts. MoS$_2$ flakes are transferred from SiO$_2$/Si substrates using PMMA and KOH solution onto the Au/SiO$_2$ substrate. Finally, metallic nanostripe-array metasurfaces were defined by e-beam lithography followed by a 50 nm Au film deposition and lift-off. The optical properties of the MoS$_2$ crystals were characterized by micro-photoluminescence (PL) spectroscopy, all the spectra were measured at room temperature (RT) using an inverted microscope coupled with an Andor spectrometer equipped with a intensified CCD camera. The excitation was provided by a temperature stabilized continuous laser (405 nm). The photocurrent was measurement using a semiconductor characterization system (Keithley 4200 SCS) under the excitation of a diode laser (650 nm).

**Conclusion**

Polarization sensitive, fast-response photodetectors are designed and demonstrated in an ultra-compact way by integrating MoS$_2$ monolayer with gap-plasmon mode metasurfaces. This structure is capable of significantly improving the photodetection characteristics of 2D materials-based photodetectors such as light absorption, polarization control, and response speed. We show that these advanced performances are controlled by the strong confinement of the electromagnetic field and the resonant coupling between the excitonic resonance and the gap plasmon resonance. We show that the integration of optical metasurfaces opens up new ways of modifying and improving the performance of 2D materials-based photodetectors for a variety of applications.

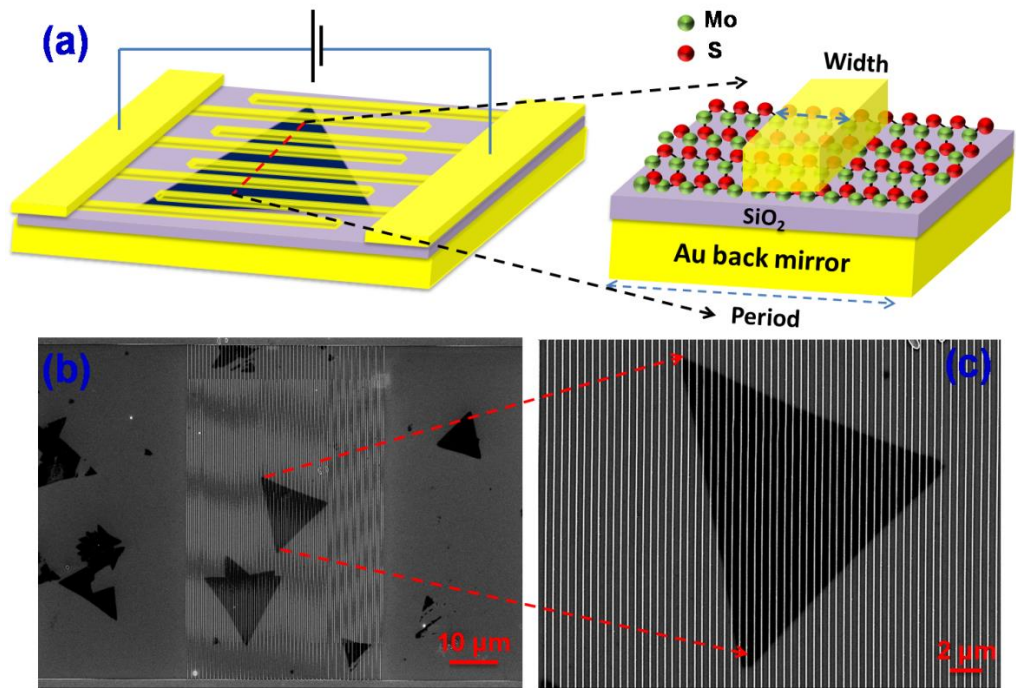

Fig. 1 The schematic (a) and the scanning electron micrograph images (b) and (c) of the fabricated hybrid $MoS_2$ gap plasmon metasurface photodetectors which comprise a monolayer $MoS_2$ flakes on the top of a 100 nm thick insulating $SiO_2$ dielectric layer sandwiched between an array of Au metallic stripe and a Au substrate. In this example, the width of the nanostripe is 75nm.

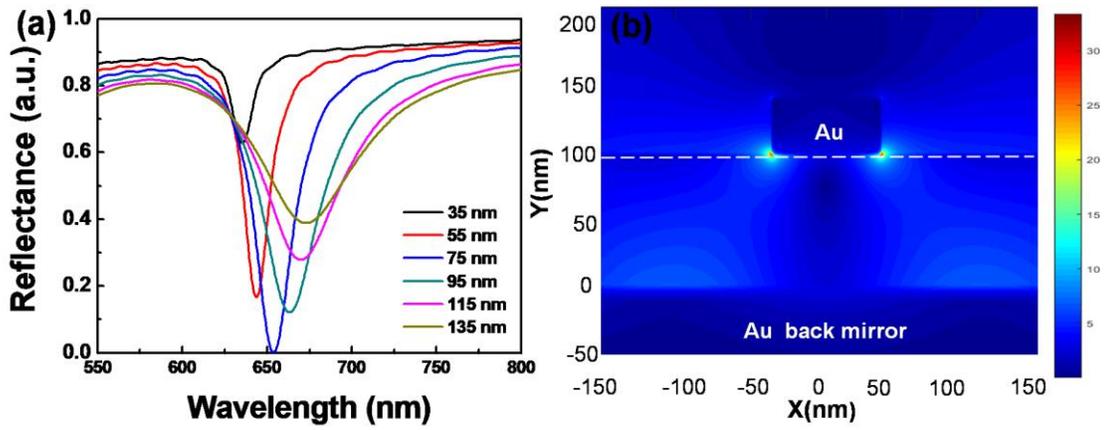

Fig. 2 The simulated reflectance of the bare gap plasmon metasurface without $MoS_2$ as a function of the width of the nanostripe (a) The calculated electric field intensity in the plane transverse to the metallic stripes showing that the electric field is tightly confined in the gap layer (b), the dash line indicates the position of active $MoS_2$ in the proposed photodetector. The period of the nanostripe array is fixed at 445 nm, i.e. sufficiently small to match roughly with the diffusion length of the photogenerated carriers.

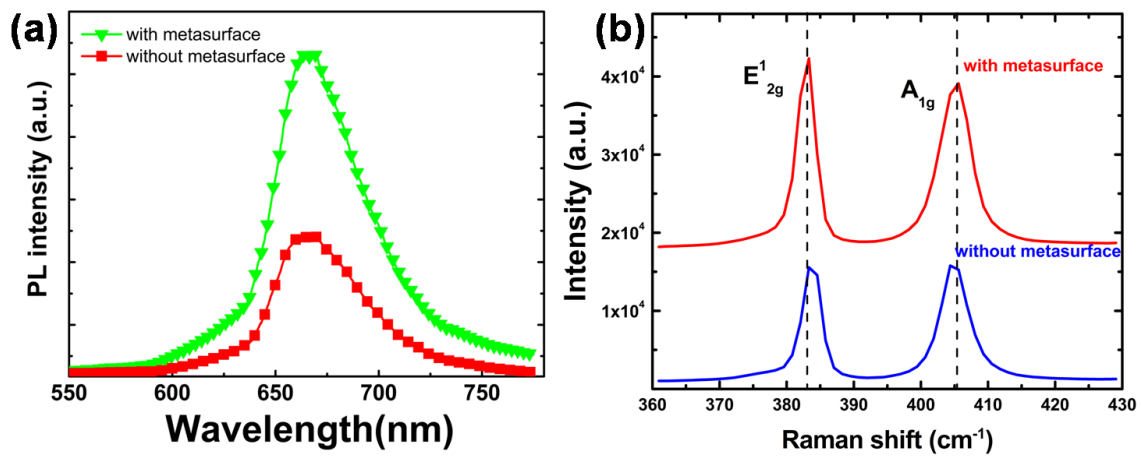

Fig. 3 (a) Photoluminescence of $MoS_2$ both with and without the GSP metasurface under the same pumping conditions. The large confinement of electromagnetic field inside the GSP metasurface leads to strong PL enhancement; (b) the Raman spectra of $MoS_2$ monolayer inside and outside the plasmonic cavity show only slight difference likely due to the induced strain. Note that the A1g peak position remains unchanged, indicating no charge transfer from the plasmonic cavity to the underlying $MoS_2$ flakes

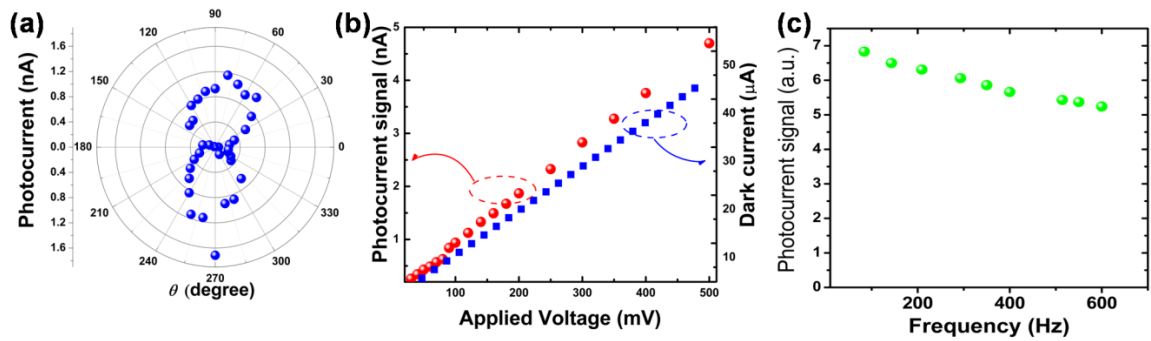

Fig. 4 (a) the photocurrents measured at different polarization angles in polar coordinates demonstrate a large polarization sensitivity of the proposed hybrid photodetector structure; (b) the measured photocurrent and the dark current as a function of applied voltage exhibit a linear dependence on the applied voltage, revealing the photoconductive characteristic of the proposed structure; (c) the photocurrent slightly decreases when increasing the switching frequency of incident light, indicating its capacity of fast response for photodetection applications.

Acknowledgements and Funding:

We acknowledge financial support from the European Union's Horizon 2020 under the European Research Council (ERC) grant agreement No. 639109 (project Flatlight). ADLB acknowledges financial support from SEP-CONACYT Ciencia Basica grant No 258674 and CONACYT-ERC grant 291826. X. Yang acknowledges financial support from Natural Science Foundation of China (61804136) and China postdoctoral science foundation (2018M630829).